\documentclass[a4paper]{jpconf}
\usepackage{graphicx}

\newcommand{\bequ}{\begin{equation}}
\newcommand{\eequ}{\end{equation}}
\newcommand{\bea}{\begin{eqnarray}}
\newcommand{\eea}{\end{eqnarray}}

\newcommand{\tev}{\, {\rm TeV}}
\newcommand{\gev}{\, {\rm GeV}}

\begin{document}
\title{Lepton Flavour Violating Decays\\ in the Littlest Higgs Model with T-Parity}

\author{Cecilia Tarantino}

\address{Physik Department,
Technische Universit\"at M\"unchen,
D-85748 Garching, Germany}

\ead{tarantino@fis.uniroma3.it}

\begin{abstract}
We present the results of an extensive analysis of lepton flavour violating decays in the Littlest Higgs model with T-parity (LHT).
As lepton flavour violation is highly suppressed in the
Standard Model by small neutrino masses, the LHT effects turn out to
be naturally huge and could be seen in the near future experiments.
\end{abstract}

\section{The LHT Model}
The {\it little hierarchy problem}, i.e. the problem of hierarchy between
a low ($\approx 10^2\gev$) Higgs mass and a quite high ($>10\tev$) Standard
Model (SM) cutoff scale indicated by electroweak (ew) precision measurements,
has been one of the main motivations to elaborate models of New Physics (NP).
While Supersymmetry is at present the leading candidate, 
different proposals have been formulated more recently.
Among them, Little Higgs models play an important role, being
perturbatively computable up to about $10 \tev$ and with a rather small number 
of parameters.

In Little Higgs models\cite{ACG} the Higgs is naturally light as it is
identified with a Nambu-Goldstone 
boson of a spontaneously broken global
symmetry, whose gauge and Yukawa interactions are incorporated 
without generating quadratic one-loop mass corrections, through the
so-called {\it collective symmetry breaking} (CSB). 
Indeed, the CSB has the peculiarity of generating
the Higgs mass only when two or more couplings in the Lagrangian are
non-vanishing, thus avoiding one-loop quadratic divergences.
Diagrammatically, the CSB is realized through the contributions of new 
particles with masses around $1 \tev$, that cancel the SM quadratic
divergences.

The most economical, in matter content, Little Higgs model is the Littlest
Higgs (LH)\cite{ACKN}, where the global group $SU(5)$ is spontaneously broken
into $SO(5)$ at the scale $f \approx \mathcal{O}(1 \tev)$ and
the SM ew sector is embedded in an $SU(5)/SO(5)$ non-linear sigma model. 
Gauge and Yukawa Higgs interactions are introduced by gauging the subgroup of
$SU(5)$: $[SU(2) \times U(1)]_1 \times [SU(2) \times U(1)]_2$. 
In the LH model, the new particles appearing at the $\tev$ scales are the heavy
gauge bosons ($W^\pm_H, Z_H, A_H$), the heavy top ($T$) and the scalar triplet 
$\Phi$.

In the LH model, however, ew precision tests are satisfied only for quite large
values of the NP scale, $f \ge 2-3 \tev$\cite{HLMW,CHKMT}, due to
tree-level heavy gauge boson contributions and the triplet vacuum 
expectation value (vev). 
The LH model can be reconciled with ew precision tests by
introducing a discrete symmetry called T-parity\cite{CL}, which
acts as an automorphism that exchanges the $[SU(2) \times U(1)]_1$ 
and $[SU(2) \times U(1)]_2$ gauge factors. 
As T-parity explicitly forbids the tree-level contributions of  heavy gauge
bosons and the interactions that induced the triplet vev,
the compatibility with ew precision data can be obtained already for smaller 
values of the NP scale, $f \ge 500 \gev$\cite{HMNP}.
Another important consequence is that particle fields are T-even or T-odd
under T-parity. The SM particles and the heavy top
$T_+$ are T-even, while the heavy gauge bosons $W_H^\pm,Z_H,A_H$ and the
scalar triplet $\Phi$ are T-odd.
Additional T-odd particles are required by T-parity: 
the odd heavy top $T_-$ and the so-called mirror fermions, i.e.,
fermions corresponding to the SM ones but with opposite T-parity and $\mathcal{O}(1 \tev)$ mass.
Mirror fermions are characterized by new flavour interactions with SM fermions
and heavy gauge bosons, which involve two new unitary 
mixing
matrices in the quark sector, $V_{Hd}$ and
$V_{Hu}$ satisfying $V_{Hu}^\dagger V_{Hd}=V_{CKM}$, and two in the lepton
sector, $V_{H\ell}$ and $V_{H\nu}$ satisfying $V_{H\nu}^\dagger V_{H\ell}=V_{PMNS}^\dagger$\cite{HLP,SHORT}. 

Because of these new mixing matrices, the Littlest Higgs model with T-parity
(LHT) does not belong to the Minimal
Flavour Violation (MFV) class of models\cite{UUT,AMGIISST} and significant 
effects in flavour observables are possible.
Other LHT peculiarities are the rather small number of new particles and
parameters (the SB scale $f$, the parameter $x_L$ describing $T_+$ mass and
interactions, the mirror fermion masses and $V_{Hd}$ and $V_{H\ell}$
parameters) and the
absence of new operators in addition to the SM ones.
On the other hand, one has to recall that Little Higgs models are low
energy non-linear sigma models, whose unknown UV-completion introduces a
theoretical uncertainty reflected by a left-over logarithmic cut-off dependence\cite{BPUB,BBPRTUW} in $\Delta F=1$ processes.

\section{Lepton Flavour Violation in the LHT Model}
Several studies of flavour physics have been performed in the LHT model in the 
last three years, for both quark\cite{HLP,BBPRTUW,BBPTUW} and lepton sectors\cite{Indian,BBDPT}.
They show that the LHT mirror fermion interactions can yield large NP effects
in the quark sector, mainly in $K$ and $B$ rare and CP-violating
decays\cite{BBPRTUW}, and that even larger NP effects are possible in
the lepton sector\cite{Indian,BBDPT}. 
The smallness of ordinary
neutrino masses, in fact, assures that the mirror fermion contributions to
lepton flavour violating (LFV) decays represent by far the dominant effects.

In\cite{BBDPT} we have studied the most interesting LFV processes: $\ell_i
\rightarrow \ell_j \gamma$, $\tau \rightarrow \ell P$ (with $P=\pi, \eta, \eta'$), $\mu^- \rightarrow e^-
e^+ e^-$, the six three-body decays $\tau^- \rightarrow l_i^- l_j^+ l_k^-$ and 
the rate for $\mu-e$ conversion in nuclei.
We have also calculated the rates for $K_{L,S} \rightarrow \mu e$, $K_{L,S}
\rightarrow \pi^0 \mu e$, $B_{d,s} \rightarrow \mu e$, $B_{d,s} \rightarrow
\tau e$ and $B_{d,s} \rightarrow \tau \mu$.

The number of significant experimental constraints on flavour violating decays
is rather limited in the lepton sector .
Basically only the upper bounds on $Br(\mu \rightarrow e \gamma)$\cite{muegamma}, $Br(\mu^-
\rightarrow e^- e^+ e^-)$\cite{meee}, $Br(K_L \rightarrow \mu e)$\cite{KLmue-exp} and $R(\mu Ti
\to e Ti)$\cite{mue-conv_bound} can be used in our analysis.
The situation may change significantly in the coming years thanks to near
future experiments\cite{mue-conv_bound,megexp,SuperB,J-PARK}.
Meanwhile, we have estimated the LHT effects, imposing the experimental bounds 
mentioned above and scanning over mirror lepton masses in the range $[300
\gev, 1500 \gev]$ and over the parameters of the $V_{H\ell}$ mixing matrix,
with the symmetry breaking scale $f$ fixed to $f=1 \tev$ or $f=500 \gev$ 
in accordance with ew precision tests\cite{HMNP}.
We note that for $f=500 \gev$ also the very recent experimental upper bounds on
$\tau \to \mu \pi, e \pi$ given in \cite{Banerjee}, where
Belle\cite{Belle-radiative,Belle-semi} and BaBar\cite{teg-exp,BaBar-semi} results have been combined, become effective.

We have found that essentially all the rates considered can reach or approach
present experimental upper bounds\cite{BBDPT}.
In particular, in order to suppress the $\mu \rightarrow e \gamma$ and $\mu^-
\rightarrow e^- e^+ e^-$ decay rates below the experimental upper bounds, the
$V_{H\ell}$ mixing matrix has to be rather hierarchical, unless mirror
leptons are quasi-degenerate. 

Moreover, following the strategy proposed
in\cite{Ellis:2002fe,Arganda:2005ji,Paradisi1} in the supersymmetric framework, we have identified certain correlations between
branching ratios that are less parameter dependent than the individual branching ratios
and could provide a clear signature of the model.
In particular, we find that the ratios $Br(\ell_i \to \ell_j \ell_j
\ell_j)/Br(\ell_i \to \ell_j \gamma)$, $Br(\ell_i \to \ell_j \ell_j
\ell_j)/Br(\ell_i \to \ell_j \ell_k \ell_k)$ and $Br(\ell_i \to \ell_j \ell_k
\ell_k)/Br(\ell_i \to \ell_j \gamma)$ could allow for a transparent
distinction between the LHT model and the MSSM (see Table~\ref{tab:comparison}).
\begin{table}
{\renewcommand{\arraystretch}{1.5}
\begin{center}
\begin{tabular}{|c|c|c|c|}
\hline
ratio & LHT  & MSSM (dipole) & MSSM (Higgs) \\\hline\hline
$\frac{Br(\mu^-\to e^-e^+e^-)}{Br(\mu\to e\gamma)}$  & \hspace{.8cm} 0.4\dots2.5\hspace{.8cm}  & $\sim6\cdot10^{-3}$ &$\sim6\cdot10^{-3}$  \\
$\frac{Br(\tau^-\to e^-e^+e^-)}{Br(\tau\to e\gamma)}$   & 0.4\dots2.3     &$\sim1\cdot10^{-2}$ & ${\sim1\cdot10^{-2}}$\\
$\frac{Br(\tau^-\to \mu^-\mu^+\mu^-)}{Br(\tau\to \mu\gamma)}$  &0.4\dots2.3     &$\sim2\cdot10^{-3}$ & $0.06\dots0.1$ \\
$\frac{Br(\tau^-\to e^-\mu^+\mu^-)}{Br(\tau\to e\gamma)}$  & 0.3\dots1.6     &$\sim2\cdot10^{-3}$ & $0.02\dots0.04$ \\
$\frac{Br(\tau^-\to \mu^-e^+e^-)}{Br(\tau\to \mu\gamma)}$  & 0.3\dots1.6    &$\sim1\cdot10^{-2}$ & ${\sim1\cdot10^{-2}}$\\
$\frac{Br(\tau^-\to e^-e^+e^-)}{Br(\tau^-\to e^-\mu^+\mu^-)}$     & 1.3\dots1.7   &$\sim5$ & 0.3\dots0.5\\
$\frac{Br(\tau^-\to \mu^-\mu^+\mu^-)}{Br(\tau^-\to \mu^-e^+e^-)}$   & 1.2\dots1.6    &$\sim0.2$ & 5\dots10 \\
$\frac{R(\mu Ti\to e Ti)}{Br(\mu\to e\gamma)}$  & $10^{-2}\dots 10^2$     & $\sim 5\cdot 10^{-3}$ & $0.08\dots0.15$ \\\hline
\end{tabular}
\end{center}\renewcommand{\arraystretch}{1.0}
}
\vspace*{-0.5cm}
\caption{\it Comparison of various ratios of branching ratios in the LHT model
  and in the MSSM without and with significant Higgs
  contributions.}
\label{tab:comparison}
\end{table}

Finally, we have studied the muon anomalous magnetic moment finding that,
even for values of the NP scale $f$ as low as $500 \gev$, $a_\mu^{LHT}<1.2\cdot 10^{-10}$.
This value is roughly a factor $5$ below the current experimental
uncertainty\cite{Bennett:2006fi}, implying that the possible discrepancy between the SM prediction
and the data cannot be solved in the LHT model. 

\vspace*{0.2cm}
I would like to thank the other authors of the work presented here:
Monika Blanke, Andrzej J. Buras, Bj\"orn Duling and  Anton Poschenrieder.

\end{document}